# MS-DCANet: A Novel Segmentation Network For Multi-Modality COVID-19 Medical Images

Xiaoyu Pan[1,*], Huazheng Zhu[2,*], Jinglong Du[1], Guangtao Hu[1], Baoru Han[1], Yuanyuan Jia[1]

[1]College of Medical Informatics, Chongqing Medical University, Chongqing, People's Republic of China; [2]College of Intelligent Technology and Engineering, Chongqing University of Science and Technology, Chongqing, People's Republic of China

*These authors contributed equally to this work

Correspondence: Yuanyuan Jia, Email yyjia@cqu.edu.cn

**Aim:** The Coronavirus Disease 2019 (COVID-19) pandemic has increased the public health burden and brought profound disaster to humans. For the particularity of the COVID-19 medical images with blurred boundaries, low contrast and different infection sites, some researchers have improved the accuracy by adding more complexity. Also, they overlook the complexity of lesions, which hinder their ability to capture the relationship between segmentation sites and the background, as well as the edge contours and global context. However, increasing the computational complexity, parameters and inference speed is unfavorable for model transfer from laboratory to clinic. A perfect segmentation network needs to balance the above three factors completely. To solve the above issues, this paper propose a symmetric automatic segmentation framework named MS-DCANet. We introduce Tokenized MLP block, a novel attention scheme that uses a shift-window mechanism to conditionally fuse local and global features to get more continuous boundaries and spatial positioning capabilities. It has greater understanding of irregular lesion contours. MS-DCANet also uses several Dual Channel blocks and a Res-ASPP block to improve the ability to recognize small targets. On multi-modality COVID-19 tasks, MS-DCANet achieved state-of-the-art performance compared with other baselines. It can well trade off the accuracy and complexity. To prove the strong generalization ability of our proposed model, we apply it to other tasks (ISIC 2018 and BAA) and achieve satisfactory results.

**Patients:** The X-ray dataset from Qatar University which contains 3379 cases for light, normal and heavy COVID-19 lung infection. The CT dataset contains the scans of 10 patient cases with COVID-19, a total of 1562 CT axial slices. The BAA dataset is obtained from the hospital and includes 387 original images. The ISIC 2018 dataset is from the International Skin Imaging Collaborative (ISIC) containing 2594 original images.

**Results:** The proposed MS-DCANet achieved evaluation metrics (MIOU) of 73.86, 97.26, 89.54, and 79.54 on the four datasets, respectively, far exceeding other current state-of-the art baselines.

**Conclusion:** The proposed MS-DCANet can help clinicians to automate the diagnosis of COVID-19 patients with different symptoms.

## Introduction

As one of the most severe epidemics in human history, COVID-19 continues to threaten human's life and health. By the end of December 2022, the World Health Organization said 645,084,824 cases of COVID-19 had been diagnosed, of which 6,633,118 cases died. Even though RT-PCT (Reverse Transcription Polymerase Chain Reaction) testing is internationally considered the gold standard for COVID-19 screening,[1] it may return a high rate of false negatives. Radiological imaging has also been used as an important complement to RT-PCR and shown itself to be effective, accurate and reliable.

X-ray and CT (Computed Tomography) are two typical radiological imaging techniques for COVID-19 diagnosis, follow-up assessment and evaluation of disease progress. X-ray is usually used as a cost-effective preliminary test for COVID-19 infection, while CT scanning exhibits a higher diagnostic rate for COVID-19 and enables clear visualization of lung lesions. However, with the growing number of suspected COVID-19 cases, manual segmentation of lung infections becomes time-consuming, and the resulting markings are subjective in nature.

Improved AI (artificial intelligence) pushes DL (Deep Learning) to a new stage.[2] Currently, most mainstream automatic segmentation methods rely on a symmetric Encoder-Decoder U-shaped framework, and UNet with its extension methods have become the de-facto medical image choice. These models based on FCNNs (Fully convolutional neural networks), show remarkable performance in extracting local features, making them highly effective in various medical image segmentation tasks. Important improvement frameworks based on FCNN include UNet,[3] UNet++,[4] UNet3+,[5] Attention-UNet,[6] ResU-Net++,[7] DC-UNet,[8] and some others. For example, the encoder of FCNN-based networks[3,9] uses a module that combined convolutional layers and down-sampling layers to increase the depth of network in order to enlarge the receptive field. In the decoder, skip connections bridge the high-resolution and low-resolution features, compensating for the absence of pixel values in the down-sampling layer and providing precise location information. This mechanism ensures accurate segmentation results.

Recently, Transformer-based architectures with a local-to-global attention mechanism have achieved state-of-the-art results in multiple medical image segmentation challenge datasets. Stimulated by the great success of Transformer in NLP (Natural Language Processing), Dosovitskiy et al[10] proposed the application of ViT (Vision Transformer) to computer vision and it has been fully demonstrated on image classification tasks. To improve the generalization capability of ViT, Liu et al[11] replaced the MSA (Multi-headed Self-attention) module in traditional Transformer with an SW-MSA (Shifted Windows Multi-Head Self-Attention) module to combine multiple Swin-Transformer blocks into one Swin-Transformer. Restricting message communication to adjacent windows has improved target detection and semantic segmentation ability. In recent years, numerous Transformer-based U-shape architectures, including TransUNet,[12] Swin-UNet,[13] SUNet[14] among others, have been fully applied to medical image segmentation tasks. Transformer may be seen as a CNN inheritor, borrowing from CNN such good inspirations as inductive bias, residual connection, etc. To solve the feature loss problem of CNN-based tasks, Transformer could retain more spatial information by extracting global contexts. Transformer also has better migration capabilities and tends to perform better in downstream tasks with large datasets.

In contrast to the CNN and Transformer frameworks, MLP (Multilayer Perceptron) models have emerged as a noteworthy alternative. These models learn all parameters from the MLP's linear layer, resulting in remarkable results that are similar to CNN and Transformer models. More research are dedicated to further exploring this area of study.[15,16]

Although various improved UNet frameworks have been proposed to detect and segment medical images, the application of existing methods to COVID-19 infection segmentation remains challenging.[17,18] Both COVID-19 infections can vary significantly in size and shape, apart from which most COVID-19 medical images show blurred boundaries, dense noise points, low contrast, and significant variation of organ shape in different periods.

Appropriately fusing multi-scale features enhances network generalization and self-adaptation. CA-Net[19] improves adaptability by introducing three hybrid attention mechanisms for multi-scale images. Based on ConvNeXt, ConvUNeXt[20] designs a lightweight attention mechanism to focus on target regions. MS-RED[21] enhances contextual and local feature recognition through the fusion of multi-scale residual features. While these methods use various attention mechanisms to fuse multi-scale features, they fail to consider the impact of spatial feature correlation in the process of segmentation and ignore the location correlation among multiple target objects. This oversight may result in incorrect positioning and the absence of crucial features.

Inspired by current research on U-shape networks, we have introduced a novel MLP-based framework specifically designed for high-efficient segmentation of COVID-19 infections. The work can be summarized as follows:

1. We propose MS-DCANet, a symmetric segmentation network with a multi-level feature fusion and novel attention mechanism.
2. We use the Tokenized MLP block to perform conditional fusion of local and global features to get more continuous boundaries and spatial positioning capabilities. Also, we improve the recognize irregular small targets ability by adding DC-blocks and Res-ASPP modules.
3. We fully validate its generalization ability on a multi-modality COVID-19 task and two other tasks (BAA and ISIC 2018) and both get a positive feedback.
4. MS-DCANet can achieve a well trade-off between accuracy, complexity and inference speed. Also, we introduce a more lightweight framework named MS-DCANet-S while still achieving the satisfied segmentation results.

## Related Works
In this section, we will briefly introduce the existing methods, from different technical aspects. A thorough comparison between our purposed MS-DCANet and other COVID-19 Infection Segmentation network will be detailly argued the advantages of the MS-DCANet.

### MLP-Based Approaches
MLPs have received attention in CV and NLP for their own natural attention mechanisms. However, their memory and computation-intensive requirements become a key bottleneck for practical applications. Tolstikhin et al[15] discarded the traditional attention and convolution modules in order to propose the MLP-Mixer framework, which adopted both channel mixing MLP and token mixing MLP layers, the two kinds of MLP layers being executed alternately to exchange information on row and column dimensions and achieving a segmentation effect similar to ViT on JFT-300M and ImageNet datasets. It successfully extracts image features without employing convolution, thanks to its unique channel interaction mechanism. But its generalisation capability cannot be further improved by its large number of parameters, making it challenging to extend it to downstream tasks. Lian et al[16] designed an AS-MLP framework based on the initial MLP Mixer framework with axial shifts to focus on local information by adding a shift window, extracting features from the vertical and horizontal dimensions, and then combining features from these two dimensions to achieve information exchange and fusion. Based on the axial shifts mechanism, it can capture more local features by using multi-directional information. Furthermore, it has a greatly decreased number of parameters comparing to the MLP-Mixer framework and has been applied to target detection and semantic segmentation tasks successfully. Experiments have shown that direct summation of vertical and horizontal features prone to large amounts of information redundancy and limit global modelling capabilities. Similar to AS-MLP, Valanarasu et al[22] incorporated tokenized MLP blocks into UNet and established the UNeXt framework with a symmetric Encoder-Decoder. This framework extracted local-to-global semantic information corresponding with different axial shifts by executing sequential feature transformations on Width and Height dimensions for the input feature maps. It is worth noting that feature transformations are not simply feature summation in two dimensions in the manner of AS-MLP. Compared with AS-MLP, UNeXt can improve the spatial exploitation of information. Lacking the necessary parameters, its poor performance on small target object segmentation cannot be ignored.

### COVID-19 Infection Segmentation
While automatic segmentation technology based on AI has great potential for segmentation of COVID-19 infections, it also faces challenges. Problems limiting the development of automatic segmentation technology include blurred edge of infected focus, low contrast between infected tissue and normal tissue, and lack of a large number of training images. Fan et al[23] proposed Inf-Net, it can aggregate high-level features to obtain contextual information to guide segmentation. Also, it focuses on building relationships between regions and boundaries in order to obtain clear boundaries. It proposes a semi-supervised framework that can be adapted to few-shot learning. Wang et al[24] proposed COPLE-Net to mitigate down-sampling's information loss problem by combining max-pooling and average pooling, using the bridge layer to mitigate the semantic gap between encoder and decoder. Zhou et al[25] introduced a automatic network that effectively reduces complexity while addressing the challenge of detecting small targets in large scenes. Additionally, it shows potential in segmenting COVID-19 lesions. Also seeking to reduce the model's complexity, Qiu et al[26] proposed MiniSeg, a lightweight network that is not easily over-fitted, has high computational efficiency and is easy to deploy. Karthik et al[27] chose a pixel-level attention model to obtain CT contour boundaries and used these features to refine areas infected by COVID-19. Although various algorithms have improved the accuracy of COVID-19 infections segmentation, many existing networks overlook the complexity of lesions, hindering their ability to effectively capture the relationship between segmentation sites and the background, as well as the edge contours and global context. MS-DCANet addresses this limitation by conditional fusion of local and global features to get more continuous boundaries and spatial positioning capabilities. Furthermore, MS-DCANet shows competitive performance in handling small target objects with irregular boundaries.

# Methods

## Network Overview

MS-DCANet (Multi-Scale Dual Channel Attention Network) is a symmetric MLP-based network comprising four Tokenized MLP blocks, four DC-blocks (Dual-Channel blocks), and a bottleneck Res-ASPP (Resnet Atrous Spatial Pyramid Pooling)), as shown in Figure 1. In the encoder, we use DC-blocks to obtain multi-level features, while the Tokenized MLP block projects the features into tokens and uses axial shifts to transform different dimensions to obtain more contextual information. The decoder utilizes residual blocks to recover local features lost during encoding. It achieves this by merging low-resolution features from the encoder with high-resolution features from the decoder through a skip connection. It is worth noting that except for using Tokenized MLP blocks to precisely locate key spatial locations, we also introduce an Attention Gate mechanism before the DC-block to suppress skip connections in non-related regions, helping to retain more high-level semantic information and obtain more refined segmentation. In Figure 1, C1, C2, C3, C4 and C5 represent each layer's channel number and we test the effect of different channel numbers on the model's performance. We also propose MS-DCANet-S, a more light-weight framework obtained by reducing the number of filters in each layer. MS-DCANet-S greatly reduces computational complexity and parameters while maintaining segmentation accuracy across four tasks.

## DC-Block with Attention Gate

Aiming at the problem of fuzzy objects and background interference in COVID-19 medical images, we present DC-block with an Attention Gate structure to obtain a greater proportion of multi-scale features in order correctly to separate background from ROI (Region of Interest). The Attention Gate can improve the accuracy of segmented objects by suppressing skip connections in non-related regions. As shown in Figure 2, the Attention Gate contains two parts of the input; G is the semantic information from the decoder below, and X passes the semantic information from the encoder to the decoder through the skip connections. The final result is the input of DC-block. To obtain more semantic features, DC-block adopts a two-channel sequence of 3 × 3 convolutional layers to replace the single-layer convolution and combines the semantic features of the convolution layer of dual channels to get the output. The three 3 × 3 filters are W/ 6, W/3, and W/2, respectively, and W is the number of filters in each layer. In order to avoiding overfitting, batch normalization operation is performed before each layer's Activation Function (RELU).

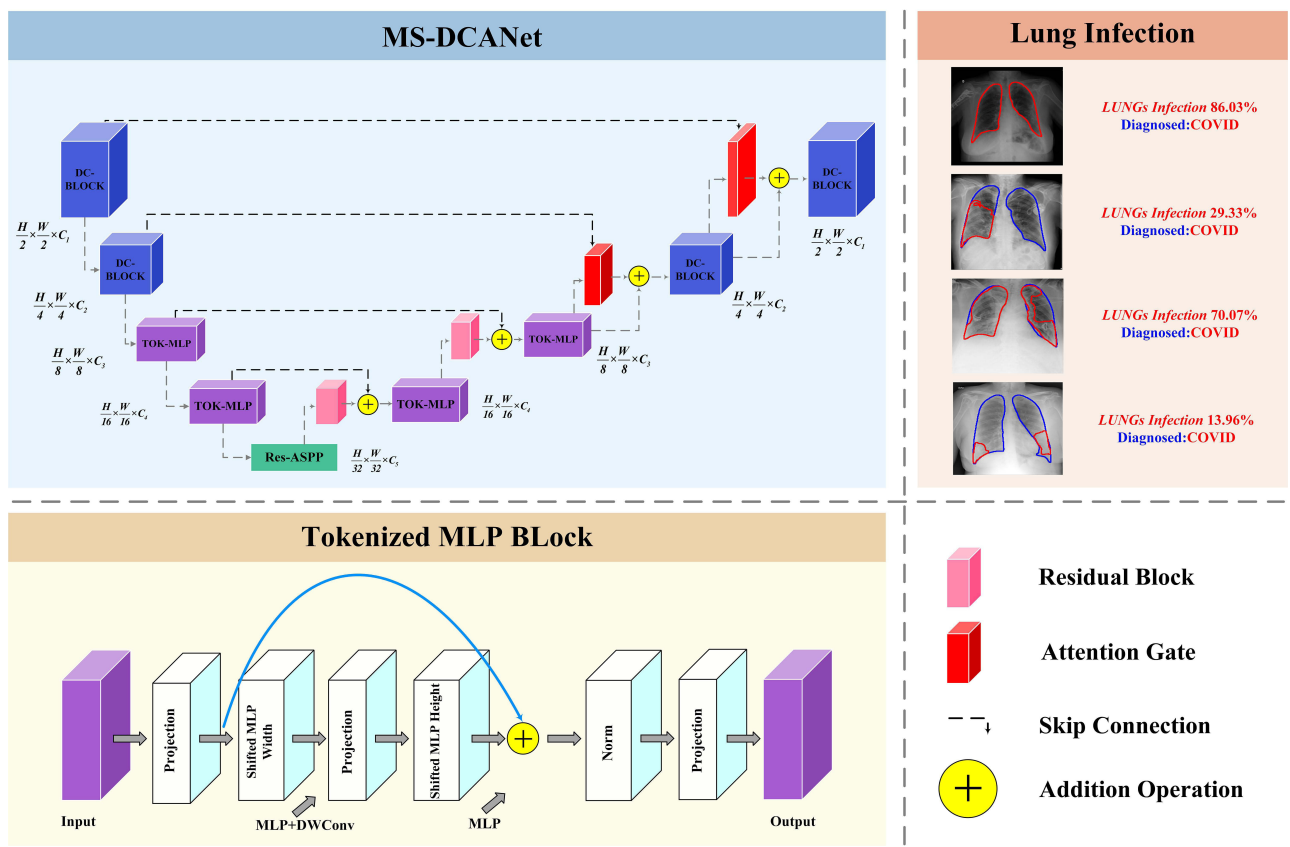

**Figure 1** The architecture of MS-DCANet.

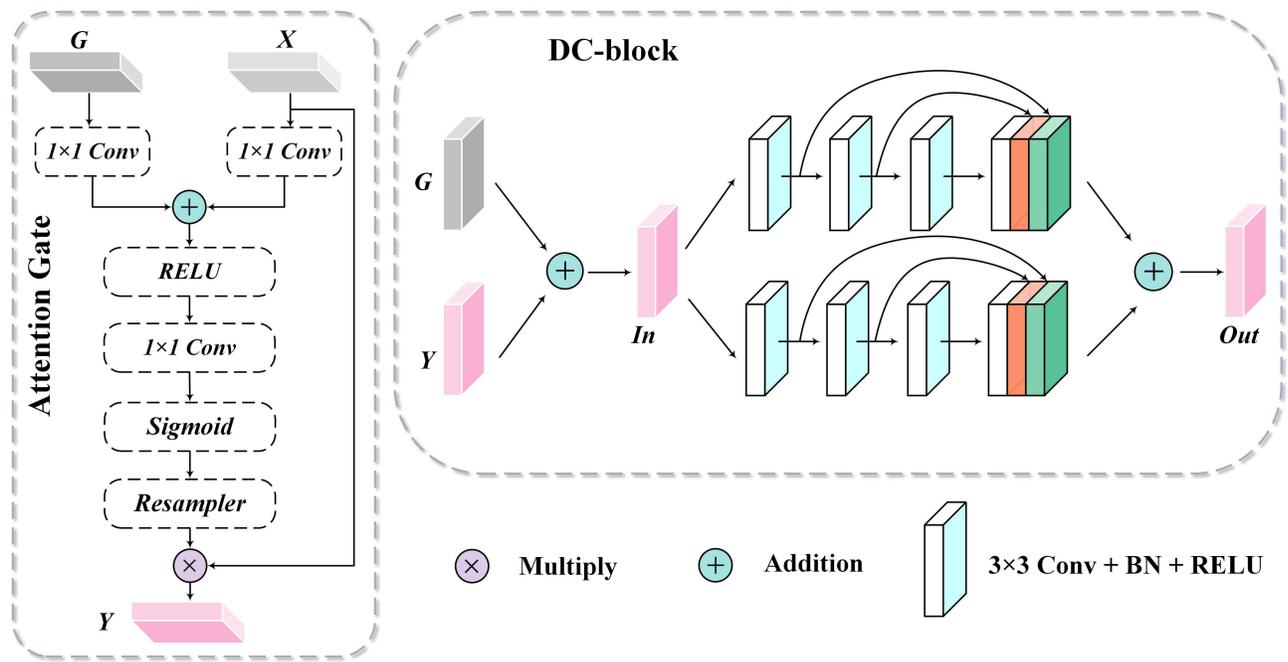

**Figure 2** DC-block and Attention Gate architecture in the decoder.

## Tokenized MLP Block

Tolstikhin et al[15] proposed MLP-Mixer to improve the model's performance and at the same time reduce the number of parameters and computational complexity. Unlike a transformer-based structure, MLP-Mixer can use linear layers to gain global dependency similar to self-attention, and it works well on ImageNet, JFT-300M. The framework is illustrated in Figure 3a. An MLP contains two fully connected layers and an activation function (GELU). However, not all pixel points require the use of long-range dependencies in practical tasks and, in such situations, MLP-Mixer will cause a large number of low-level features to be missed. To solve these limitations, Lian et al[16] introduced the AS-MLP structure with axial shifts based on the MLP-Mixer, and pioneered its use for such downstream tasks as target detection and semantic segmentation, ultimately achieving positive feedback. Figure 3b shows the AS-MLP structure, including normalized layer, axial shift and MLP. In axial shift, all features were first projected onto a linear layer through channel projection before being shifted simultaneously in both vertical and horizontal dimensions and then reunited in both dimensions after channel projection. Finally, axial shift projected the features again by a channel projection layer. To attain semantic dependency from local to global, AS-MLP realized the flow of information in different spatial positions by merging the features of two dimensions. While AS-MLP has significant advantages over MLP-Mixer in extraction of local features, it still cannot fully realize the long-range dependence of features, and more research based on MLP is underway.

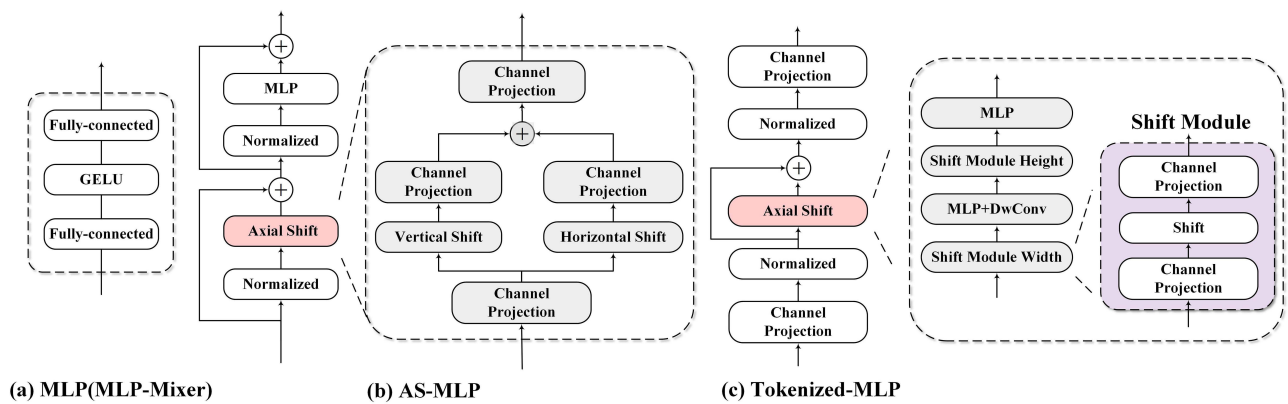

**Figure 3** Compare the structures of two Axial Shifts.
**Note:** Data from[15,16]

To solve the above issues, this paper utilizes Tokenized MLP block, of which Figure 3c illustrates the specific architecture. Tokenized MLP block starts with channel projection to project all the features into the tokens (similar to AS-MLP), and then uses the axial shift mechanism to project the features in the width and height directions. Unlike AS-MLP which projects features in both directions at the same time, there is an order of dimension projection in axial shift. The token is first shifted in the width direction, and then in the height direction. The shift operation contributes to making the MLP focus only on regions of higher correlation and suppresses learning about features of regions of lower correlation, which enables Tokenized MLP block to learn more low-level semantics. The formulas for performing axial shift operations in each of the two directions are given below.

We first perform a channel projection for the input features and shift them in the width direction. We then use channel projection to project the features into the token. Note that, channel projection is the same as tokenized. Where $F$ represents the input tokens after channel projection, $W$ represents shift in width.

$$T_W = Projection\ (Shift_W\ (Projection(F))) \tag{1}$$

Then, the Linear layer in MLP maps the feature to the sample space. The location information of MLP features can be encoded by DWConv while reducing the computation. GELU is used for feature extraction in DWConv, which can preserve previous useful features while abandoning features with a low degree of correlation to the task. Where DWConv represents Depthwise Separable Convolution.

$$T = DWConv\ (MLP(T_W)) \tag{2}$$

We perform a channel projection for the features. Next, the features are passed through another shifted MLP across height. The shift operation in the height dimension is the same as the operation in the width dimension. Where H represents shift in height.

$$T_H = Projection\ (Shift_H\ (Projection(T))) \tag{3}$$

We use the linear layer to map the feature in spatial domain and connect it with the previous tokens by residual connection. Residual connection uses direct mapping to avoid information loss. To alleviate the hazard from over-fitting, following this module, we add a normalization layer for batch normalization operation. After feature extraction, we transfer the features in the width direction to the height direction. Where LN represents layer normalization.

$$X = Projection\ (LN(F + MLP(T_H))) \tag{4}$$

## Bottleneck

The proposed bottleneck (Res-ASPP) acts as a bridge between encoder and decoder. Traditional ASPP mostly includes the traditional convolutional layer, the BN (Batch Normalization layer) and the RELU (Activation function). Since ResNet can reduce the number of parameters in the network while avoiding overfitting, solving the gradient disappearance and explosion issues, we use ResNet to replace the traditional convolutional layer. Figure 4 shows the Res-ASPP framework. Res-ASPP includes four ResNet blocks, each of which contains one dilated convolution layer that can enlarge the receptive field size without increasing the number of parameters. In order to learn multi-scale features, we set the dilation rate to 4, 8, 16, and 24, respectively.

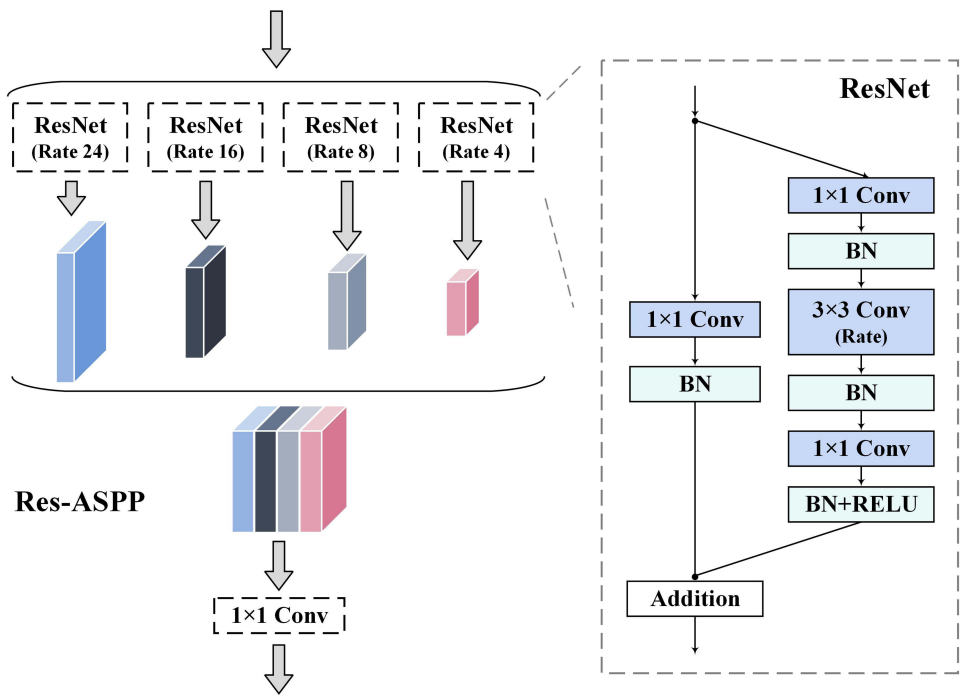

**Figure 4** The framework for Res-ASPP.

# Experiments
## Datasets

To demonstrate the segmentation accuracy of MS-DCANet for COVID-19 infections, sufficient experiments are deployed on COVID-19 tasks: COVID-19 CXR, COVID-19 CT. To further prove MS-DCANet's generalisability, we also perform experiments on two other tasks: BAA and ISIC2018. The images of each original dataset are shown in Figure 5.

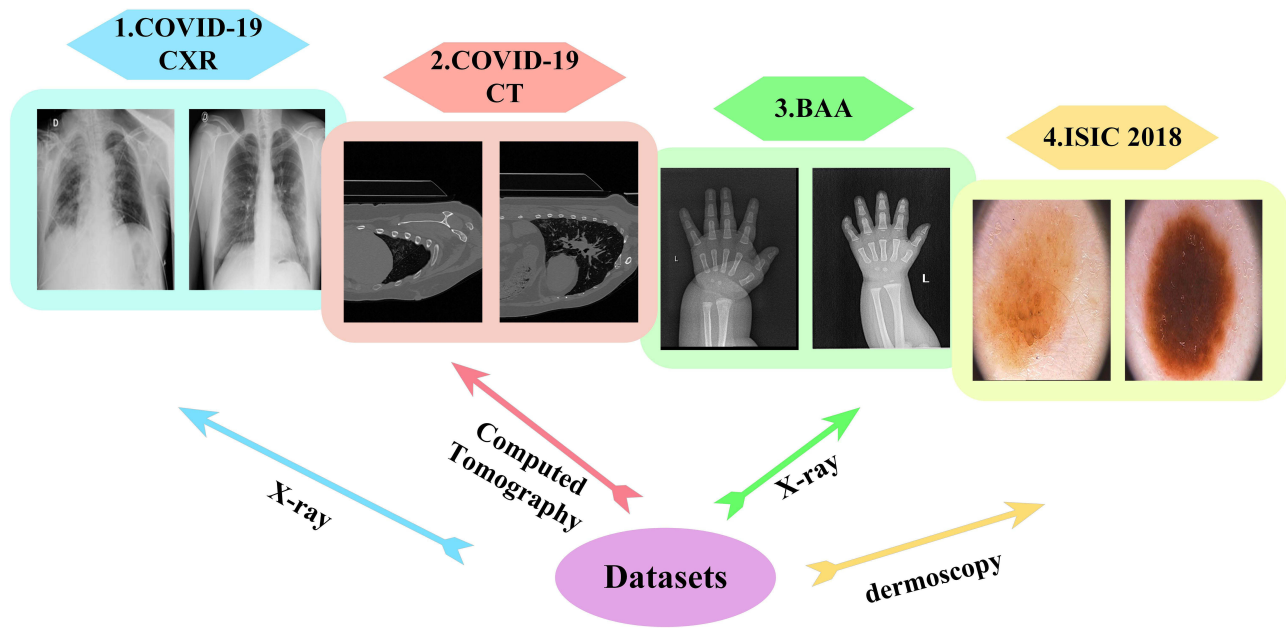

**Figure 5** The images of each original dataset.

### Covid-19 CXR

This X-ray dataset from Qatar University which contains 3379 cases with COVID-19 lung infection, we divide it into 2796 training sets and 583 testing sets for light, normal and heavy lung infection.[28–31] Our goal is to segment the infected area and calculate the percentage of the infected area for the whole lung. We adjust their size to 256×256.

### Covid-19 CT

This dataset contains CT scans of 10 patient cases with COVID-19, a total of 1562 CT axial slices, and expert segmentation labels for the lung infection. We resize each CT image to 384×384.

### BAA (Bone Age Assessment)

This dataset is obtained from the hospital and includes 387 original images from X-rays and 387 manually annotated labels. It is used to detect the age and growth of children by hand bone morphology. We select 310 of the manually annotated images as the training set and the rest of the images as the testing set. We resize each X-ray image to 512×512.

### ISIC 2018

The ISIC 2018 dataset is from the International Skin Imaging Collaborative (ISIC) containing 2594 original images and 2594 manually annotated labels.[32] In ISIC 2018, task 4 is to segment skin tumours under dermoscopy. We resize each image to 224×224.

## Evaluation Metrics

To evaluate the segmentation result in COVID-19 CXR, COVID-19 CT, BAA, and ISIC 2018 datasets, we compute F1 (balanced F Score) and MIoU (Mean Intersection over Union). To describe the segmentation metrics better, we introduce a confusion matrix to represent the prediction effect of the classifier on the test datasets. Meanwhile, we use Params Size (MB) and GFLOPs to measure the parameters and complexity of the model, respectively.

## Implementation Details

We use the pytorch framework to validate the MS-DCANet's stability. We use Adam as an optimizer to train our model. The initial learning rate is 0.0005 in four experiments, and the model's default batch setting is equal to 8 in the ISIC 2018, 4 in the BAA dataset, COVID-19 CXR datasets and COVID-19 CT. All experiments are executed on an RTX-3090 GPU with 30GB of RAM and a 7-core CPU. To solve the problem of easy overfitting in few-shot learning, in COVID-19 tasks, we only perform data augmentation (random rotation) to increase the Image and corresponding mask on the training set.

## Ablation Study

To evaluate the effect of each MS-DCANet module, we perform detailed ablation study, including 1) the impact of the target module, 2) the effect of dilation rate on Res-ASPP, 3) discuss on different numbers of Tok-MLP block, and 4) analysis on number of channels. The ablation study is done on the COVID-19 CXR dataset.

### The Impact of the Target Module

A reasonable combination of the modules can significantly improve the model's segmentation performance. Table 1 shows the configuration results between the modules. We begin with the baseline model (UNet), which shows performance improvements after introducing the Tokenized MLP block. To enlarge the receptive field while increasing network depth, we introduce bottleneck (Res-ASPP), improving the final model's predictive performance. We change the single-layer convolution block into a DC block to obtain more semantic information. The segmentation accuracy improved when we added Res-ASPP and DC-block to the Baseline. Note that simultaneously adding Res-ASPP and DC-block to the Baseline significantly improve performance with a slight increase in parameters and computation complexity. Finally, we introduce the Attention Gate (AG) module in the decoder to focus on the valuable information. Compared with the initial model version, we have reduced parameters by a considerable number, reduced computational complexity, and obtained a more robust model.

**Table 1** Ablation Study on the Target Module

| Networks | Params Size | GFLOPs | F1 | MIoU |
|---|---|---|---|---|
| Baseline (UNet) | 118.40 | 48.21 | 82.29 | 70.72 |
| Baseline+Tok-MLP | 5.61 | 0.55 | 83.23 | 71.65 |
| Baseline+Tok-MLP+Res-ASPP | 6.30 | 0.70 | 83.55 | 72.14 |
| Baseline+Tok-MLP+DC-Conv | 5.49 | 0.55 | 83.87 | 72.67 |
| Baseline+Tok-MLP+DC-Conv+Res-ASPP | 7.74 | 0.64 | 84.19 | 73.04 |
| Baseline+Tok-MLP+DC-Conv+Res-ASPP+AG | 7.90 | 0.69 | **84.65** | **73.86** |

**Note**: The bold font means the best model in this metric.

### The Effect of Dilation Rate on Res-ASPP

Dilated Convolution could obtain a multi-scale receptive field by setting different dilation rates to capture more contextual semantic information. Table 2 shows the effect of choosing different dilation rates on the model when all other configurations are the same. With dilation rates of 4, 8, 16, 24, the network achieves more remarkable performance with 0.95% and 1.89% improvement in F1 and MIoU, respectively.

### Discuss on Different Numbers of Tok-MLP Block

The above experiments show that the Tokenized MLP Block can help to capture long-distance dependencies in the medical image and refine the segmentation result. Moreover, compared with common down-sampling and up-sampling modules, the proposed method can obtain more convincing segmentation results while significantly reducing the model's parameters and computational complexity. This part will discuss the effect of different numbers of the Tok-MLP block on the segmentation effect, shown in Table 3. Setting the Tok-MLP block on the third and fourth layers gives finer segmentation with a slight parameter increase. Setting the Tok-MLP block on the second, third and fourth layer degrades segmentation performance with a slight rise in parameters compared with the second line in Table 3. Thus, we finally decided to put the Tok-MLP block in the third and fourth layers, improving F1 and MIoU by 0.80% and 1.55%, respectively.

**Table 2** Ablation Study on the Effect of Dilation Rate

| r1 | r2 | r3 | r4 | F1 | MIoU |
|---|---|---|---|---|---|
| 2 | 4 | 8 | 12 | 84.39 | 73.35 |
| 4 | 8 | 16 | 24 | **84.65** | **73.86** |
| 6 | 12 | 18 | 24 | 83.85 | 72.49 |

**Note**: The bold font means the best model in this metric.

**Table 3** Ablation Study on Different Positions of Tok-MLP

| Baseline (UNet) | Res-ASPP | Tok-MLP Layer | | | Params Size (MB) | F1 | MIoU |
|---|---|---|---|---|---|---|---|
| | | (4) | (3) | (2) | | | |
| √ | √ | √ | | | 6.17 | 83.98 | 72.73 |
| √ | √ | √ | √ | | 6.30 | **84.65** | **73.86** |
| √ | √ | √ | √ | √ | 6.44 | 84.23 | 73.15 |

**Note**: The bold font means the best model in this metric.

### Analysis on Number of Channels

Appropriately increasing the number of channels can improve the proposed network's performance. Table 4 displays the segmentation performance of small, medium, and extensive networks. C1, C2, C3, C4 and C5 are the channel numbers of each layer shown in Figure 1. Although the small network (MS-DCANet-S) is extremely lightweight, segmentation accuracy is not the best. After increasing the network channels (MS-DCANet-L), performance improves slightly but with a dramatic increase in computing cost. We finally chose the medium-sized network (MS-DCANet-M) for a better trade-off between accuracy and segmentation speed.

## Performance Comparison on COVID-19 Tasks

In the two COVID-19 datasets, we compare MS-DCANet with several state-of-the-art models: 1) UNet,[3] 2) UNet++,[4] 3) Attention U-Net.[6] We also compare several recent baselines: 1) DCSAU-Net,[33] 2) UNeXt,[22] 3) TransUNet,[12] 4) SmaAt-UNet,[34] 5) MiniSeg,[26] 6) CA-Net,[19] 7) ConvUNeXt,[20] 8) MS-RED.[21] We also use Paired *t*-test to measure the MIoU difference between the comparison model and MS-DCANet. Note that when the p-value is less than 0.05, we consider this baseline to be significantly different from our proposed MS-DCANet, where more * means a higher degree of difference. Then, we present the strength of MS-DCANet through comparing the F1-score and MIoU.

### (Tasks 1) Results on COVID-19 CXR Segmentation

We evaluate the models' performance by analyzing their parameter, computational complexity and inference speed. In practical clinical environments, the complexity of a model is crucial for successful deployment, given limitations in software and hardware resources. From Table 5, our proposed models MS-DCANet and its lightweight model MS-

**Table 4** Ablation Study on Number of Channels

| Networks | C1 | C2 | C3 | C4 | C5 | Params Size (MB) | F1 | MIoU |
|---|---|---|---|---|---|---|---|---|
| MS-DCANet-S | 8 | 16 | 32 | 64 | 128 | 1.36 | 79.67 | 67.02 |
| MS-DCANet-M | 16 | 32 | 128 | 160 | 256 | 7.90 | 84.65 | 73.86 |
| MS-DCANet-L | 32 | 64 | 128 | 256 | 512 | 21.47 | **84.87** | **74.00** |

**Note**: The bold font means the best model in this metric.

**Table 5** The Parameter Size, Computational Complexity, and FPS (Inference Speed) Differ Across Partly State-of-the-Art Baselines

| Networks | Year | Params Size (MB) | GFLOPs | FPS |
|---|---|---|---|---|
| UNet++ | 2018 | 39.89 | 26.50 | 142.54 |
| Attention U-Net | 2018 | 133.05 | 50.96 | 123.19 |
| TransUNet | 2021 | 254.88 | 24.75 | 93.17 |
| DCSAU-Net | 2022 | 9.89 | 4.88 | 60.42 |
| MiniSeg | 2021 | 0.32 | 0.13 | 67.96 |
| CA-Net | 2020 | 10.80 | 5.69 | 28.21 |
| ConvUNeXt | 2022 | 13.36 | 6.98 | 131.57 |
| MS-RED | 2022 | 15.33 | 16.75 | 37.49 |
| **MS-DCANet** | - | 7.90 | 0.53 | 112.57 |
| MS-DCANet -S | - | **1.36** | **0.11** | 127.45 |

**Note**: The bold font means the best model in this metric.

DCANet-S demonstrate notable advantages when three evaluated metrics are taken into account. Most of the models ignore the complexity of COVID-19 lesions, which lead to significant error in the procedure of actual segmentation. Our proposed model has the ability of implicit information extraction as well as edge contour detection, it can fuse multi-level features and obtain more continuous boundaries, which is well demonstrated by the follow-up experiments. Although the inference speed is slightly decreased, it is clearly acceptable compared to the huge performance benefits.

We train the MS-DCANet model on the COVID-19 CXR dataset to perform lung segmentation, and Table 6 shows the details of the experimental results. MS-DCANet achieves the best segmentation performance considering segmentation accuracy and efficiency with MIoU increases ranging from 0.48% to 10.9%. Compared with MS-RED, the benefit is practically meaningful, although not significant. On the whole, the computation complexity of MS-DCANet is satisfactory. For example, the parameter size of UNet++ is approximately five times larger than MS-DCANet, while the final segmentation result is not as good. Note that MS-DCANet is significantly different from all baselines and has a better performance in F1-score and MIoU metrics, so we can conclude that MS-DCANet is superior to all other baselines. In fact, the performance improvement of MS-DCANet is not only in the data. It can be well presented on COVID-19 lesions segmentation.

We present a few challenging COVID-19 lung images with unclear (regular and irregular) lung boundaries, high variability in lung infection features, and low contrast between infected and normal tissues. We specifically analyse the differences between baselines and our proposed model on the challenging images. Also, we compare the difference between the ratio between the predicted obtained lung infected area and the total lung area and the ratio of the real lung infected area to the total lung area. From Figure 6, it is evident that MS-DCANet excels in precisely segment lesion boundaries, specially for segmenting small objects. In contrast, the other baseline models tend to exhibit either an excessive or inadequate segmentation on challenging images. This situation is extremely obvious for UNet and UNet++. ConvUNeXt has strong space cognitive ability, and it can exactly locate multiple lesions in both lungs. However, it faces challenges when it comes to segmenting the boundaries of irregular lesions, similar to the limitations showed in MS-RED and CA-Net. In contrast, our proposed MS-DCANet can precisely segment irregular lesion boundaries, which is

**Table 6** Performance Comparison on the COVID-19 CXR Dataset (Paired *t*-test by Comparing MIoU)

| Networks | Year | F1 | MIoU | p-value |
|---|---|---|---|---|
| UNet | 2015 | 82.29 | 70.72 | *** |
| UNet++ | 2018 | 82.41 | 70.75 | *** |
| Attention U-Net | 2018 | 82.35 | 70.86 | *** |
| TransUNet | 2021 | 79.49 | 66.62 | *** |
| SmaAt-UNet | 2021 | 81.45 | 69.66 | *** |
| DCSAU-Net | 2022 | 82.93 | 71.37 | *** |
| UNeXt | 2022 | 81.46 | 69.47 | *** |
| MiniSeg | 2021 | 80.71 | 68.00 | *** |
| CA-Net | 2020 | 84.36 | 73.23 | *** |
| ConvUNeXt | 2022 | 82.88 | 71.10 | 0.167 |
| MS-RED | 2022 | 84.48 | 73.51 | 0.176 |
| **MS-DCANet** | - | **84.79** | **73.86** | - |
| MS-DCANet-S | - | 79.67 | 67.02 | *** |

**Notes**: The bold font means the best model in this metric. ***Means there is a strongly significant difference (based on MIoU).

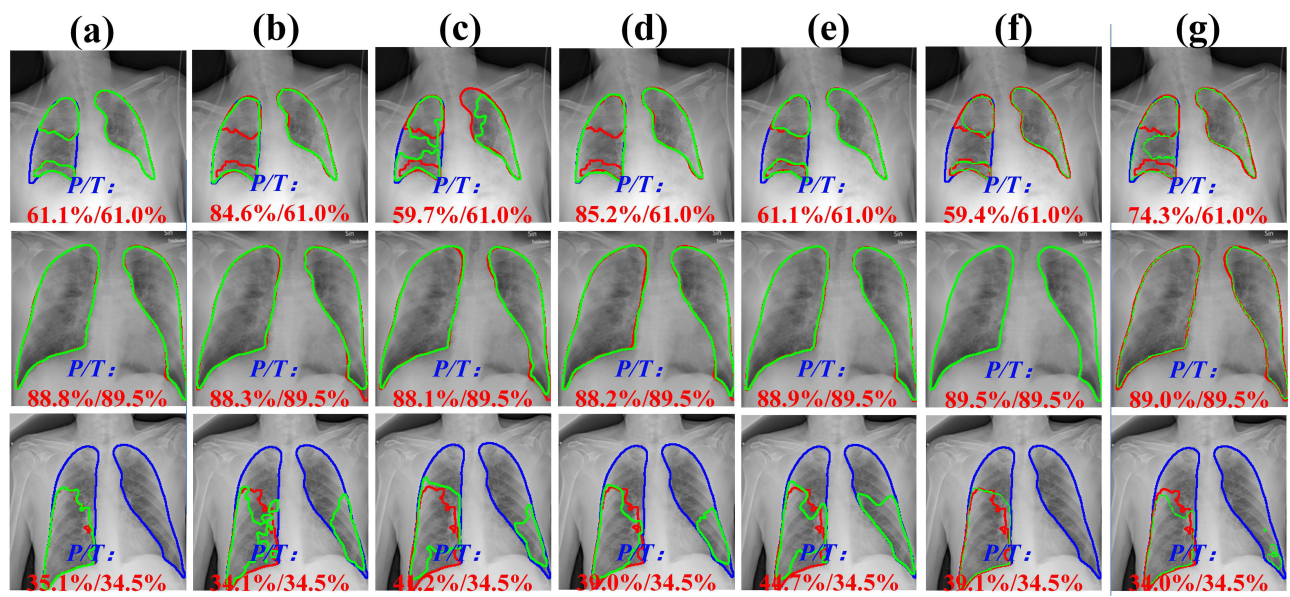

**Figure 6** The challenging COVID-19 lesions. In COVID-19 images, the green line represents the contours of the infected lesion that we predict, the red line represents the actual contours of the infected lesions, and the blue line represents the contours of the entire lung. P/T denotes the predict and actual lung infection rates, respectively. (**a**) MS-DCANet, (**b**) UNet, (**c**) UNet++, (**d**) DCSAUNet, (**e**) ConvUNeXt, (**f**) MS-RED, (**g**) CA-Net.

attributed to its global-local information exploitation and multi-level feature fusion ability. We also offer the contrast between MIoU, Params Size (MB), and GFLOPs of each comparing model in Figure 7 (COVID-19 CXR).

## (Tasks 2) Results on COVID-19 CT Segmentation

To compare the proposed model with the comparison models in COVID-19 infection segmentation from CT images, we record the experimental results in Table 7. The best results are marked in bold. From the paired *t*-test, we find that some baselines are not significantly different or weakly different from our purposed MS-DCANet, but the p-value does not fully reflect the performance of the MS-DCANet. To further assess the model's stability and robustness, we employ box plots. From Figure 8 we show that MS-DCANet has good stability, compared with other existing methods.

In Table 7, we can see that remarkable segmentation results achieved by various U-shaped networks with distinctive attention mechanisms on the COVID-19 CT Dataset. To explore the role played by MLP-based attention mechanisms in MS-DCANet further, we use Grad-CAM (Gradient-weighted Class Activation Mapping) to obtain the relationship between the image features and the class weights.[35] The experiments' results are shown in Figure 9, from which we can see that layer 3 focuses on almost all lung regions, layer 4 dramatically decreases the area of attention, and the attention mechanism's suppression effect is most obvious in layer 5. As the model layers deepens, the model's receptive field increases. And the model is more interested in the crucial regions of the image while constantly suppressing the skip-connections of irrelevant regions so as to achieve the coarse-to-fine segmentation process. Meanwhile, we use red, blue and green lines to show the model's defects with different types of attention mechanisms and our proposed model MS-DCANet, with details enlarged to fully analyse segmentation defects. We find that MS-DCANet is better than the other two models in processing COVID-19 images' contour boundaries. It makes greater use of local information, so the edge contours of the processed images are smoother. Although the other two models can capture the whole contour of the model, they do not deal well with local features, and they introduce some small features which are not related to each other. Although the MS-DCANet still has a slight bias for the overall lesion area capturing, the CRF (Conditional Random Field) and TTA (Test Time Augmentation) strategies can be adopted in the actual segmentation procedure to compensate for the initial image segmentation's shortcomings. We also offer the contrast between MIoU, Params Size (MB), and GFLOPs of each comparing model in Figure 7 (COVID-19 CT). On GFLOPs, the lightweight model MS-DCANet is roughly similar to UNeXt. As a result, the segmentation ability of the MS-DCANet has significant improvement, with a slightly increased on GFLOPs.

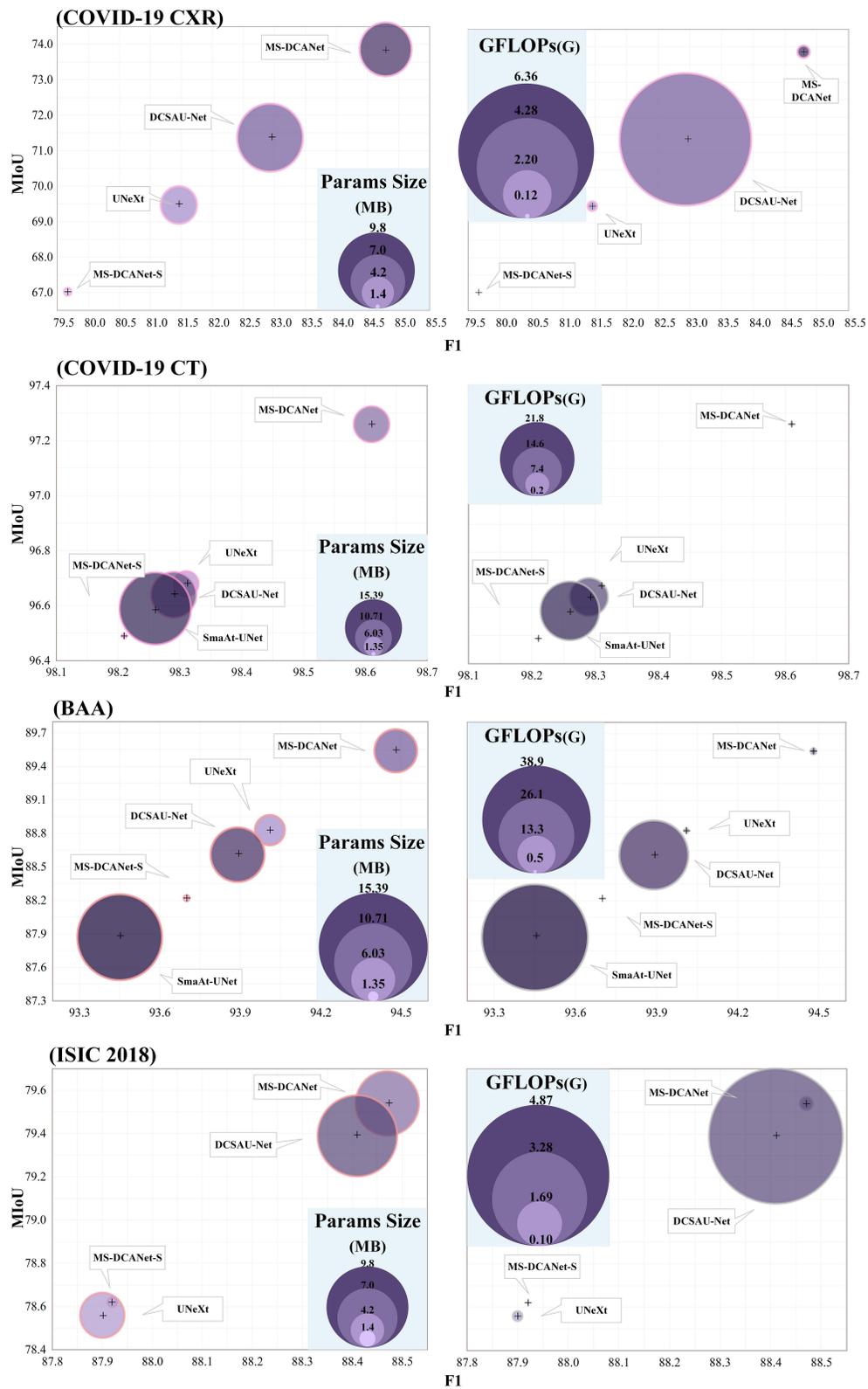

**Figure 7** The prediction results of each baseline model are compared on four datasets. Y-axis represents MIoU, X-axis represents F1, where larger values represent better model accuracy. Combining the results of the four datasets, we found that MS-DCANet can better trade-off the relationship between efficiency and accuracy than the lightweight UNeXt model and ultimately achieve better prediction results.

Table 7 Performance Comparison on the COVID-19 CT Dataset (Paired *t*-test by Comparing the MIoU)

| Networks | Year | F1 | MIoU | p-value |
|---|---|---|---|---|
| UNet | 2015 | 97.31 | 94.79 | *** |
| UNet++ | 2018 | 97.79 | 95.69 | *** |
| Attention U-Net | 2018 | 98.37 | 96.80 | *** |
| TransUNet | 2021 | 98.01 | 96.11 | 0.011*** |
| SmaAt-UNet | 2021 | 98.26 | 96.59 | 0.104 |
| DCSAU-Net | 2022 | 98.29 | 96.64 | 0.040** |
| UNeXt | 2022 | 98.31 | 96.68 | 0.069* |
| MiniSeg | 2021 | 97.91 | 95.93 | *** |
| CA-Net | 2020 | 98.50 | 97.04 | 0.100* |
| ConvUNeXt | 2022 | 98.11 | 96.31 | *** |
| MS-RED | 2022 | 98.07 | 96.23 | *** |
| **MS-DCANet** | - | **98.61** | **97.26** | - |
| MS-DCANet-S | - | 98.21 | 96.49 | 0.011** |

**Notes**: The bold font means the best model in this metric. *Means there is a difference, **Means there is a significant difference, ***Means there is a strongly significant difference (based on MIoU).

## Performance Comparison on Other Tasks

To validate MS-DCANet's generalizability, the proposed model was trained on two other tasks with the experimental procedure similar to the COVID-19.

### (Tasks 3) Generalization on BAA Datasets

Table 8 shows the results of training the proposed MS-DCANet on the BAA dataset to perform hand bone segmentation. MS-DCANet shows a remarkable improvement in segmentation accuracy, with MIoU gains ranging from 0.65% to 3.96%. In particular, a more clear hand-bone boundary can be segmented after applying the multi-scale semantic fusion approach to the UNeXt. This is due to MS-DCANet's ability to diminish information loss after down-sampling compressed images, resulting in more delicate low-level cues. MS-DCANet also yields good segmentation results with 3.96% and 2.14% improvement in MIoU and F1 compared with the transformer-based model TransUNet. Finally, even though UNet++ has clear advantages in small datasets, it cannot fully trade-off accuracy and efficiency. It has achieved competitive segmentation results based on huge parameters, which remarkably decrease its efficiency. It is evident that MS-DCANet has a significant advantage comparing with the other CNN-based and Transformer-based models on the BAA dataset.

Figure 10 visualizes the segmentation effect of MS-DCANet and several typical UNet models on the BAA dataset. The red boxes highlight the regions where MS-DCANet can do better than the other models. Most models have over-segmentation and under-segmentation in the small sample dataset because they lose too much important information. MS-DCANet performs better than the other models because it obtains accurate position information so that it has more contiguous segmentation boundaries. The results show that learning multi-scale features in MS-DCANet enhances segmentation accuracy. We also offer the comparison between MIoU, Params Size (MB), and GFLOPs of each models in Figure 7 (BAA). The performance of MS-DCANet shows an 0.8% improvement under the same parameters and operation complexity as UNeXt. We show that MS-DCANet can better trade-off accuracy and operational complexity.

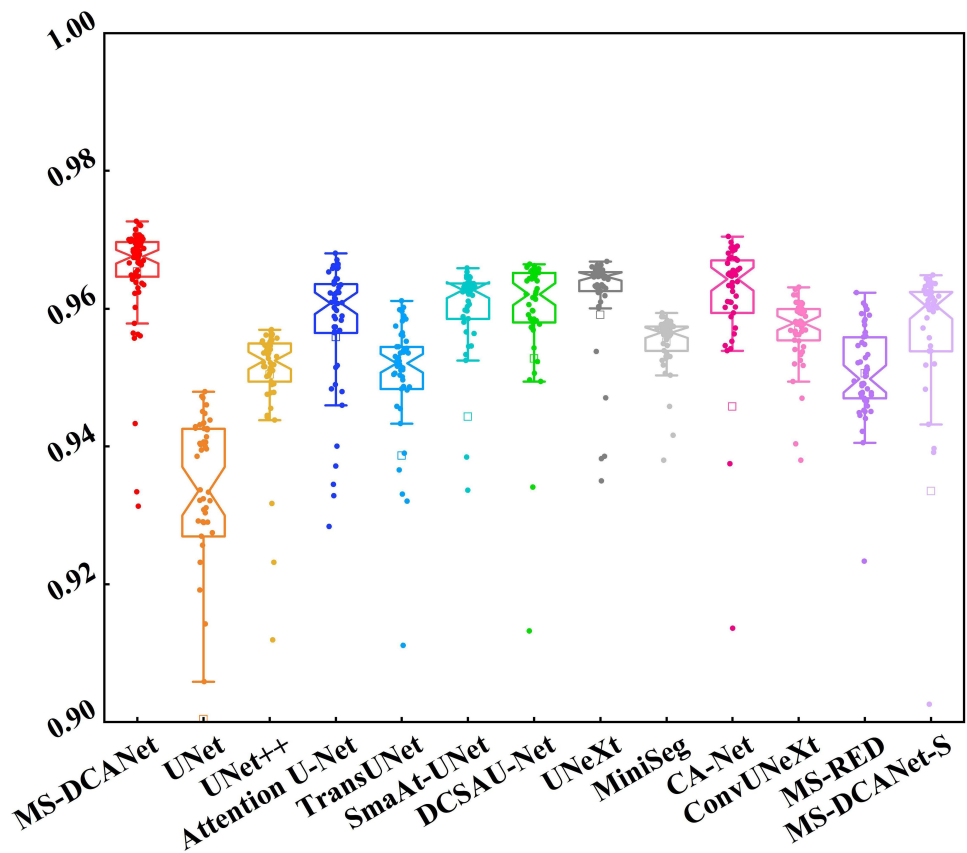

**Figure 8** Stability and robustness analysis of each model, based on MIoU.

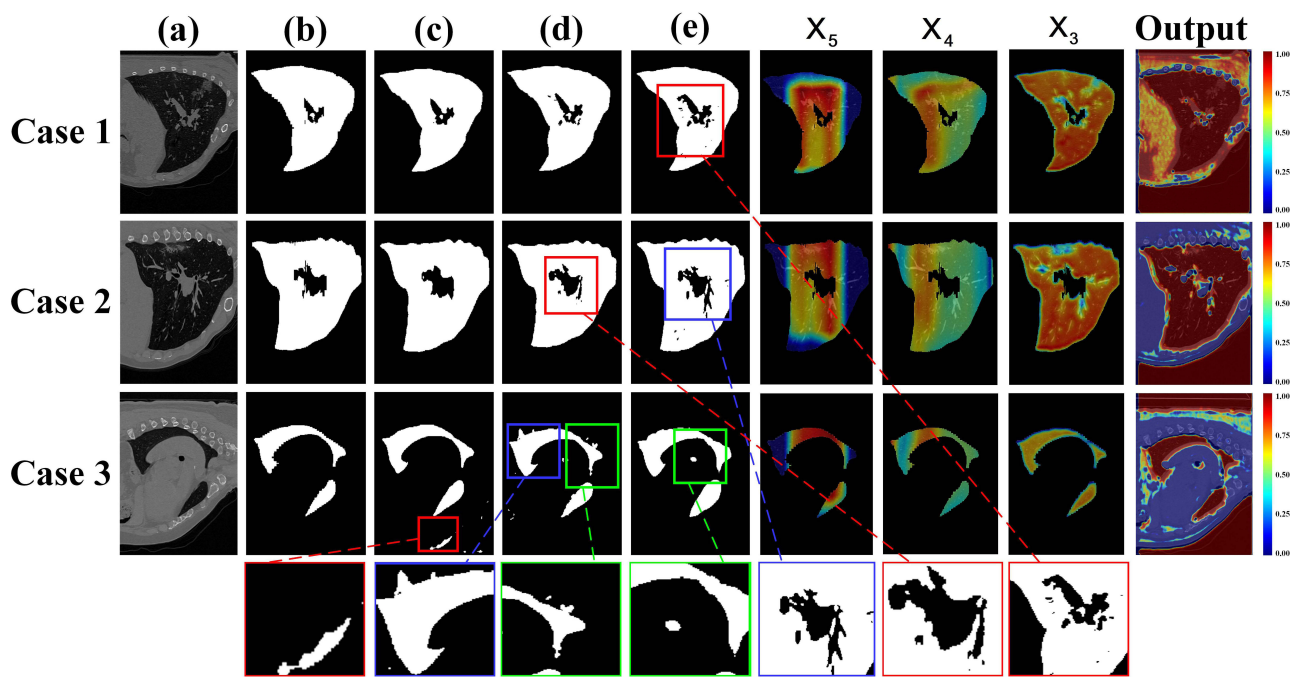

**Figure 9** To explore the role of attentional mechanisms on COVID-19 CT Dataset. Where F3 represents layer 3, F4 represents layer 4, F5 represents layer 5. (**a**) Origin, (**b**) Ground Truth, (**c**) MS-DCANet, (**d**) SmaAt-UNet, (**e**) Attention U-Net.

**Table 8** Performance Comparison on the BAA Dataset (Paired *t*-test by Comparing the MIoU)

| Networks | Year | F1 | MIoU | p-value |
|---|---|---|---|---|
| UNet | 2015 | 92.68 | 86.51 | *** |
| UNet++ | 2018 | 94.01 | 88.82 | *** |
| Attention U-Net | 2018 | 92.96 | 86.90 | *** |
| TransUNet | 2021 | 92.50 | 86.13 | *** |
| SmaAt-UNet | 2021 | 93.45 | 87.87 | *** |
| DCSAU-Net | 2022 | 93.89 | 88.61 | *** |
| UNeXt | 2022 | 94.01 | 88.83 | 0.263 |
| CA-Net | 2020 | 94.12 | 88.96 | *** |
| ConvUNeXt | 2022 | 92.65 | 86.41 | *** |
| **MS-DCANet** | - | **94.48** | **89.54** | - |
| MS-DCANet-S | - | 93.70 | 88.22 | *** |

**Notes**: The bold font means the best model in this metric. ***Means there is a strongly significant difference (based on MIoU).

In clinical practice, obtaining enough clinical samples to train the network is difficult due to the existence of privacy policies and the diversity of disease types. Medical images also suffer from blurred lesion boundaries, noise, and artifacts. Thus, we decided to exert different types of Gaussian and Poisson noise for high-resolution images to generate low-resolution medical images and assess the model's robustness by comparing the segmentation effects of different models on low-resolution images. Figure 11 shows the simulated low-resolution images.

To examine the proposed model's segmentation accuracy on low-resolution images, we evaluate the segmentation quality of the images by adding Sensitivity, Precision, and ASD (the Average Surface Distance) metrics based on the comparison of F1 and MIoU of the model. Higher values of sensitivity and precision mean better segmentation results, as does a smaller ASD value. From Figure 12, we can see that MS-DCANet significantly outperforms UNet in all metrics except Sensitivity in low-resolution image with Poisson noise. This is because MS-DCANet can obtain more local information and get the spatial position information accurately, so it performs better on unclear images. It also produces excellent performance on the severely damaged low-resolution image (Gaussian Noise (0.45)). Thus, we observe that MS-DCANet has good robustness and stability.

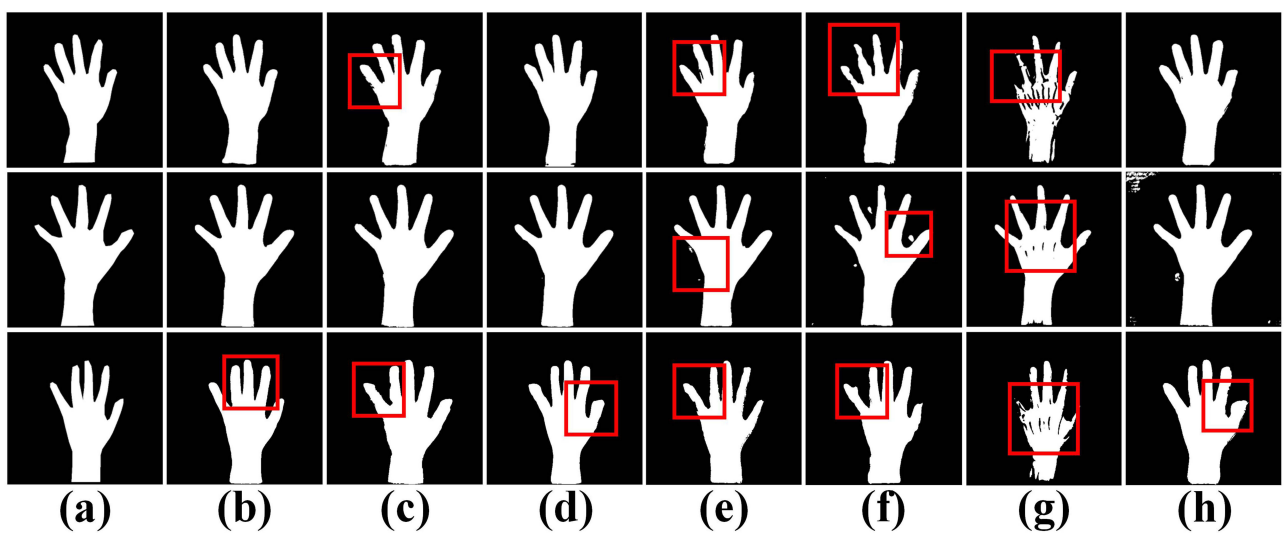

**Figure 10** Comparison of segmentation performance of several models on the BAA dataset. (**a**) Ground Truth, (**b**) MS-DCANet, (**c**) UNet, (**d**) UNet++, (**e**) Attention U-Net, (**f**) MS-DCANet-S, (**g**) UNeXt, (**h**) DCSAU-Net.

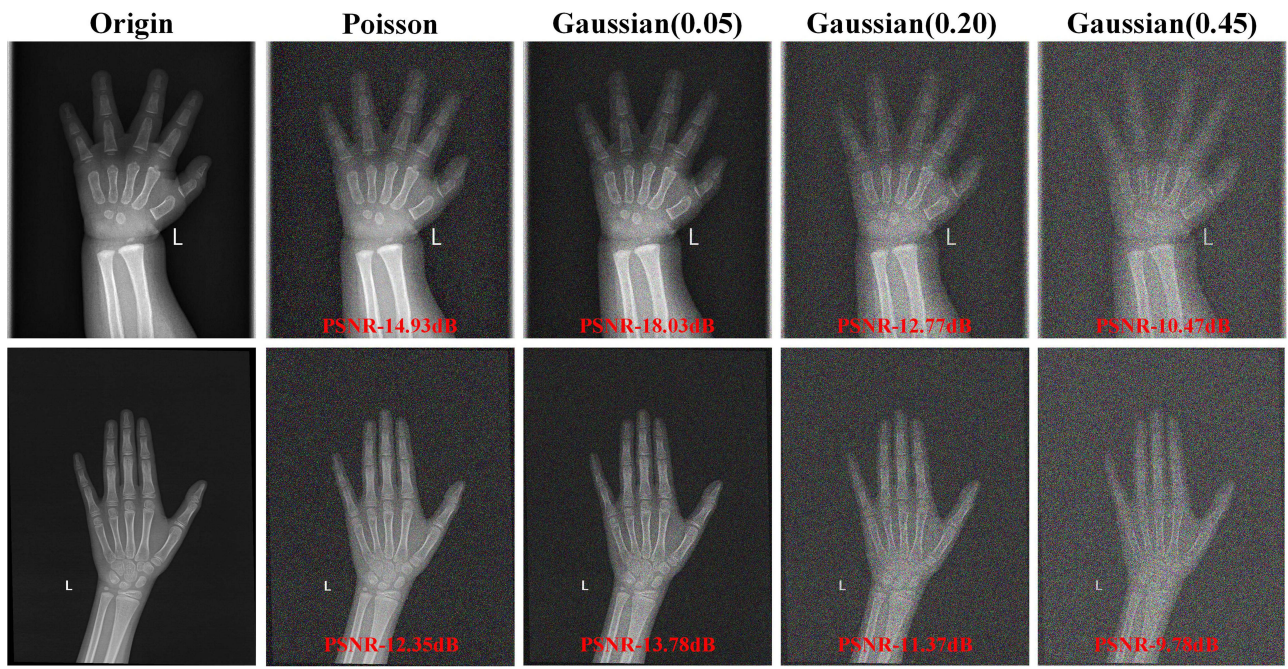

**Figure 11** The simulated noisy low-resolution images of the BAA dataset.

### (Tasks 4) Generalization on ISIC 2018 Datasets

Finally, we further test the proposed model's generalisability on ISIC 2018 dataset and record the experimental results in Table 9. MS-DCANet achieves the best segmentation results of all the models, and the model performance is significantly

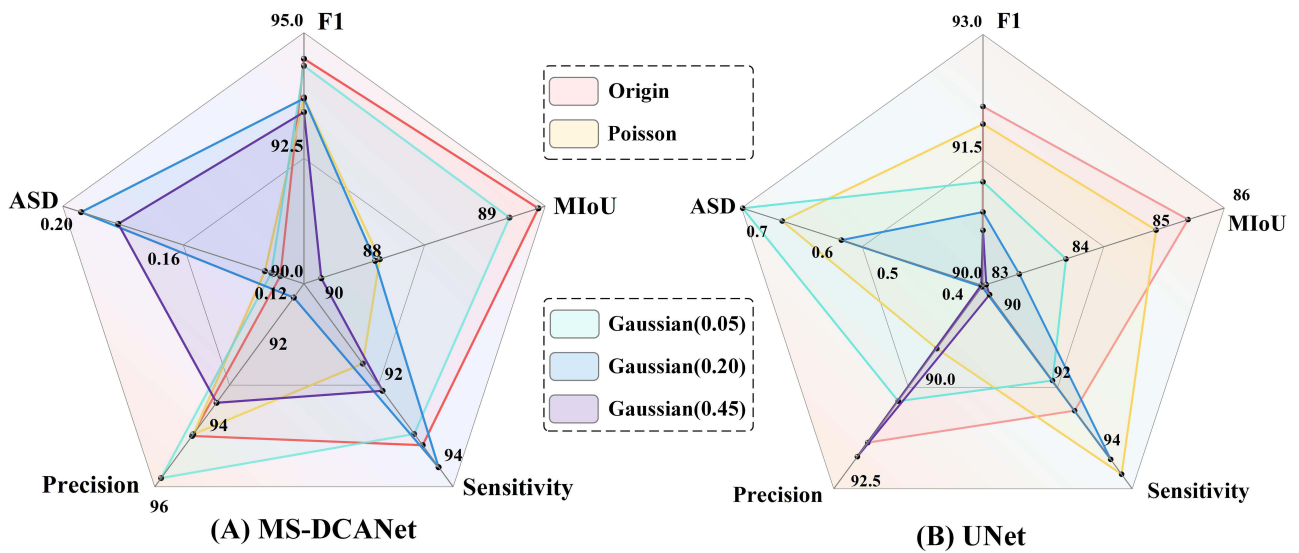

**Figure 12** The model's segmentation accuracy for low-resolution images under a different type of noise. We add evaluation metrics to fully evaluate the impact of different levels of noise images on the experiment. Poisson denotes Poisson noise, Gaussian(s) denotes Gaussian noise, and s is the variance. The larger the variance, the stronger the noise and the lower the quality of the processed image. Except for the ASD, the smaller the better, all remaining metrics are the larger the better.

**Table 9** Performance Comparison on the ISIC 2018 Dataset (Paired *t*-test by Comparing the MIoU)

| Networks | Year | F1 | MIoU | p-value |
|---|---|---|---|---|
| UNet | 2015 | 86.64 | 76.66 | *** |
| UNet++ | 2018 | 87.89 | 78.54 | *** |
| Attention U-Net | 2018 | 87.51 | 77.91 | *** |
| TransUNet | 2021 | 86.59 | 76.52 | *** |
| Swin-UNet | 2021 | 86.38 | 76.17 | *** |
| DCSAU-Net | 2022 | 88.41 | 79.39 | *** |
| UNeXt | 2022 | 87.90 | 78.56 | *** |
| CA-Net | 2020 | 88.24 | 79.13 | *** |
| ConvUNeXt | 2022 | 87.00 | 77.26 | *** |
| **MS-DCANet** | - | **88.47** | **79.54** | - |
| MS-DCANet -S | - | 87.92 | 78.62 | *** |

**Notes**: The bold font means the best model in this metric. ***Means there is a strongly significant difference (based on MIoU).

improved with similar parameters and complexity as UNeXt. It also achieves good results with its lighter weight structure, which balances accuracy and complexity. We reduce the Params Size by 32 times while decreasing the computational complexity by 47 times and the model segmentation performance (MIoU) rises by 3.95% when compared with TransUNet. Based on Paired *t*-test, all baselines differ from MS-DCANet. Meanwhile, MS-DCANet's F1-score and MIoU are better than other baselines, indicating that MS-DCANet can achieve state-of-the-art performance on the ISIC 2018 dataset.

We select several images of skin tumours with challenging segmentation in ISIC 2018, and the segmentation effects of different models are shown in Figure 13. DCSAU-Net can obtain an approximately correct segmentation contour but easily over-segments and does not correctly understand the information at the edge of the lesion. TransUNet and Swin-UNet did not show the segmentation advantage in small sample datasets, while UNet and UNet++ did not handle the contextual semantic information correctly, resulting in severe loss of location information. MS-DCANet achieves good segmentation results from challenging datasets, and the lesion edges are clearly segmented. We also offer the comparison between MIoU, Params Size (MB), and GFLOPs of each model in Figure 7 (ISIC 2018). Compared with other lightweight models DCSAU-Net, the parameters of MS-DCANet were reduced by 25%, but can achieve better segmentation results.

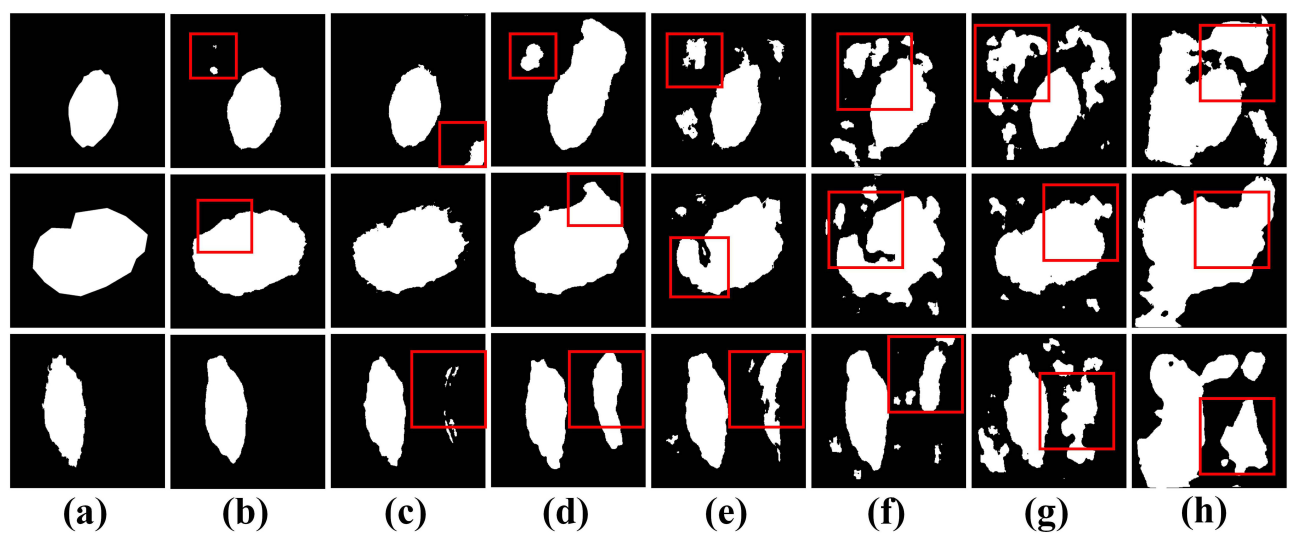

**Figure 13** Comparison of segmentation performance of several baseline models on the ISIC 2018 dataset. (**a**) Ground Truth, (**b**) MS-DCANet, (**c**) DCSAU-Net, (**d**) MS-DCANet-S, (**e**) Attention U-Net, (**f**) UNet, (**g**) UNet++, (**h**) UNeXt.

## Conclusion

In this paper, we detailly analyse the characteristic of COVID-19 infections and the advantages and disadvantages of classical CNN-based and Transformer-based models. We discover that the classical UNeXt model is too lightweight leading to blurred segmentation boundaries due to excessive semantic gaps in the actual segmentation process. To overcome these problems, we have designed a symmetric Encoder-Decoder MS-DCANet model leveraging DC-block and Res-ASPP block to obtain Multi-Scale features. Tokenized MLP block is used to get a self-attention mechanism similar to the transformer, which makes the MS-DCANet model extract semantic dependency on local-to-global. We have performed full validation on multi-modality COVID-19 dataset to measure the segmentation performance of the MS-DCANet. All datasets contain challenging images such as medical images with low contrast, a large number of artifacts and ill-defined boundaries. MS-DCANet can outperform the other state-of-the-art U-shape networks on these challenging images. The segmentation effect is significantly enhanced with a slight increase in number of parameters and operational complexity. MS-DCANet can better trade-off accuracy and complexity. We also test MS-DCANet on the other two segmentation tasks, and MS-DCANet both obtain satisfactory segmentation accuracy. Thus, we can judge that MS-DCANet has extremely high generalizability and robustness. We will continue to refine the MS-DCANet model and extend it to smaller target lesions to achieve more precise localization and segmentation.

## Ethical Approval

Issued by the Health Commission, Ministry of Education, Ministry of Science and Technology, Bureau of Traditional Chinese Medicine dated 18th Feb 2023. In this study, we validate the performance of our purposed MS-DCANet on four public open source datasets. The study did not involve human subjects and did not create additional risks or discomforts due to the research involvement. As a result, ethical approval was not required for the study.

## Author Contributions

All authors made a significant contribution to the work reported, whether that is in the conception, study design, execution, acquisition of data, analysis and interpretation, or in all these areas; took part in drafting, revising or critically reviewing the article; gave final approval of the version to be published; have agreed on the journal to which the article has been submitted; and agree to be accountable for all aspects of the work.

## Funding

This work was supported by Young Project of Science and Technology Research Program of Chongqing Education Commission of China (No.KJQN202001513 and KJQN202101501), the Natural Science Foundation of Chongqing, China (Grant cstc2021jcyi-bshX0168), medical image intelligent analysis and application innovation team of Chongqing medical university (NO.ZSK0102), the Wise Information Technology of Medical Project of Chongqing Medical University (ZHYXQNRC202205).

## Disclosure

Professor Guangtao Hu reports grants from the Wise Information Technology of Medical Project of Chongqing Medical University, during the conduct of the study. The authors declare that they have no other competing interests.